\documentstyle[12pt]{article}
\newcommand{\scite}{~\cite}
\renewcommand{\baselinestretch}{1.1}
\def\beq{\begin{equation}}
\def\eeq{\end{equation}}
\def\beqa{\begin{eqnarray}}
\def\eeqa{\end{eqnarray}}
\def\rd{{\mathrm d}}
\def\pt{p_\bot}
\def\jpsi{{J\!/\!\psi}}
\def\as{\alpha_s}

\def\bentarrow{\:\raisebox{1.1ex}{\rlap{$\vert$}}\!\rightarrow}
\def\dk#1#2#3{
\begin{array}{r c l}
#1 & \rightarrow & #2 \\
 & & \bentarrow #3
\end{array}
}

\begin{document}

\begin{titlepage}

\begin{flushright}
\begin{tabular}{l}
		FNT/T-94/13 \\
            LNF-94/024(P)\\
		hep-ph/9405241 \\
            May 1994\\
\end{tabular}
\end{flushright}
\vspace{2.5cm}

\begin{center}
{\huge $\jpsi$ Production via Fragmentation at the Tevatron} \\
\vspace{2.5cm}
{\large Matteo Cacciari$^{a}$ and Mario Greco$^{b}$} \\
\vspace{.5cm}
{\sl $^a$Dipartimento di Fisica Nucleare e Teorica,
Universit\`a di Pavia, Pavia, Italy \\
INFN, Sezione di Pavia, Pavia, Italy\\
\vspace{.2cm}
$^b$Dipartimento di Fisica,
Universit\`a de L'Aquila, L'Aquila, Italy \\
INFN, Laboratori Nazionali di Frascati, Frascati, Italy }\\
\vspace{2.5cm}
\begin{abstract}
The production of $\jpsi$ at large transverse momenta ($\pt > M_\jpsi$)
in $p\bar p$ collisions is considered by including the mechanism of
fragmentation.  Both contributions of fragmentation to
$\jpsi$ and of fragmentation to $\chi$ states followed by radiative decay to
$\jpsi$ are taken into account. The latter is found to be dominant and larger
than direct production. The overall theoretical estimate is shown to be
nearly consistent with the experimental observation.
\end{abstract}

\end{center}
\renewcommand{\baselinestretch}{1.}
\vfill
\hrule
\small
\vspace{-.5cm}
\begin{tabbing}
E-mail addresses: \= cacciari@pv.infn.it \\
		  \> greco@lnf.infn.it
\end{tabbing}
\end{titlepage}

The study of the properties of the bound states of heavy quarks plays a central
role in the understanding of Quantum Chromodynamics in that it stands on the
very border between the perturbative and non-perturbative domain. In particular
it is of key importance to
have accurate estimates of the production cross sections at large tranverse
momenta for precision tests of the theory and possible evidence of new
phenomena.

So far there has been an intensive experimental study of the $c\bar c$ $1S$
vector bound state, namely $\jpsi$, in hadron collisions both at UA1\scite{ua1}
and CDF\scite{cdf}. The results  have been compared with
theoretical calculations\scite{gms} which take into
account two different
mechanisms for $\jpsi$ production: direct charmonium production, including the
contribution from the $\chi$ states, i.e.
\beq
gg\to\jpsi\>g\qquad\dk{gg,q\bar q}{\chi g}{\jpsi\>\gamma}\qquad
\dk{qg}{\chi q}{\jpsi\>\gamma}
\eeq
and the production resulting from $B$ mesons decay
\beq
\dk{p\bar p}{bX}{B\to\jpsi\>X}
\eeq
A more recent version of this calculation, which makes use of the NLO
prediction\scite{NDE} for $b$ production, is presented in ref.~\cite{mlm}.

These calculations are in disagreement with the results form CDF, the $\jpsi$
rate observed being actually higher, by a factor of two or
more, than the predicted one\scite{cdf,mlm,huth}.

It has however recently been pointed out by E.~Braaten and
T.C.~Yuan\scite{bygluon} that at large $\pt$ an additional production mechanism
comes into play, namely  the fragmentation  of a gluon or a charm quark
into a charmonium state. While being of higher order with respect
to direct production by a power of the running coupling constant $\as$,
this mechanism becomes dominant at large $\pt$ because of a factor
$O(\pt^2/m_c^2)$ which overcomes the extra power of $\as$. The
fragmentation functions describing these processes can be calculated
perturbatively. Indeed it has been argued
in~\cite{bygluon} and subsequently shown at LO in~\cite{dfm} that $\jpsi$
production
via fragmentation will overcome the direct one (i.e. $gg\to\jpsi\>g$)
at $\pt\sim 6$-8~GeV. A similar exercise for the $\chi$ production, when the
total fragmentation probability $\int D_g^\chi$ (see below) times the
$gg\to gg$ cross
sections is compared to the direct production $gg\to\chi\>g$, reveals that
fragmentation should dominate for $\pt$ already at $\sim 2$~GeV. Since this
result is at the limit
of validity of the fragmentation function approach, we can however still expect
that the fragmentation mechanism will dominate over the direct
one at $\pt$ values as low as 5-6~GeV.

\begin{figure}[t]
\begin{center}
\begin{minipage}[t]{7cm}
\begin{center}
\parbox{6cm}{
\caption{
\label{gfrag}
\small One of the diagrams for the gluon fragmentation function at the
scale $\mu = 2m_c$.}
}
\end{center}
\end{minipage}
\begin{minipage}[t]{7cm}
\begin{center}
\parbox{6cm}{
\caption{
\label{cfrag}
\small One of the diagrams for the charm fragmentation function, at the scale
$\mu = 3m_c$.}
}
\end{center}
\end{minipage}
\end{center}
\end{figure}

In this Letter we apply these ideas to a quantitative determination of the
$\jpsi$
production rate in hadron collisions, taking also into account the production
via fragmentation processes of the $\chi$ states and
subsequent radiative decays to $\jpsi$.

To this aim the following fragmentation functions play a major role:
the gluon fragmentation to $\jpsi$\scite{bygluon}, $D_g^\jpsi$
(see fig.~\ref{gfrag});
the charm (or anticharm) fragmentation to $\jpsi$\scite{bcycharm},
$D_c^\jpsi$ (see fig.~\ref{cfrag});
the charm fragmentation to $\chi$ states\scite{chen}, $D_c^\chi$; and finally
the gluon fragmentation to $\chi$ states\scite{bychi}, $D_g^\chi$ (see
fig.~\ref{gtochi}).

\begin{figure}[t]
\begin{center}
\begin{minipage}[t]{7cm}
\begin{center}
\parbox{6cm}{
\caption{
\label{gtochi}
\small One of the diagrams for the
gluon fragmentation function to the $\chi$ states, at the scale $\mu = 2m_c$.}
}
\end{center}
\end{minipage}
\begin{minipage}[t]{7cm}
\begin{center}
\parbox{6cm}{
\caption{
\label{indg}
\small One of the  ``perturbative'' contributions to the
induced gluon fragmentation function, at the scale $\mu = 4m_c$.}
}
\end{center}
\end{minipage}
\end{center}
\end{figure}

They have been all calculated
by perturbative techniques at an initial scale of the order
of the mass of the $\jpsi$. Of course in the evaluation of the actual
cross sections they must be
evolved to the appropriate
scale\footnote{We thank Paolo Nason for having provided us with the Fortran
code for
the numerical evolution of the $D_g^\chi$ fragmentation functions.},
and one gets to the usual expression
\beq
\rd\sigma(p\bar p\to \jpsi(\pt)+X) = \sum_i\int_{z_{min}}^1\rd z
\>\rd\sigma(p\bar p\to i(\pt/z)+X,\mu)D_i^\jpsi(z,\mu)
\eeq
for the $\jpsi$ production, the sum running over $g$, $c$ and $\bar c$.
A similar formula does
hold for $\chi$ production\footnote{Note that, when performing the convolution
with the $\chi_{0,2}$ fragmentation functions, due to the singularity that
these
functions display at $x=1$ an appropriate subtraction
precedure, already introduced in ref.~\cite{cg}, must be used to ensure the
convergence of the numerical integration.}.
The cross section on the right hand side
corresponds to the inclusive production of the parton $i$, convoluted with the
appropriate structure functions (throughout this work we'll use the MRS-D0
set),
and summed over all relevant parton-parton scattering processes.
$\mu$ is the factorization scale, which we will take of order
$\mu_0 = \sqrt{\pt^2 + M_\jpsi^2}$.

The evolution of the fragmentation  functions given above obeys the usual
Altarelli-Parisi (AP) equations
\begin{equation}
\mu \frac{\partial}{\partial \mu} D_i^\jpsi(z, \mu) = \sum_j
\int^{1}_{z} \frac{dy}{y} \; P_{i \rightarrow j}(z/y, \mu)
\; D_j^\jpsi (y, \mu)
\label{onehalf}
\end{equation}
Furthermore it has been pointed out in
ref.~\cite{falk} that when one considers the whole set of the AP equations,
with
the appropriate mixings taken into account, the evolution of the $D_c^\jpsi$
will induce a gluon fragmentation function through the
splitting $g\to c\bar c$ and subsequent fragmentation of one of the quarks
into a $\jpsi$ (see fig.~\ref{indg}). This process is of order $\as^3$ but,
being enhanced by a factor $\log(\mu/M_\jpsi)$, will dominate over the
the contribution from $D_g^\jpsi$ at large $\pt$.

We present our results first using the leading order (LO) expressions for the
partonic cross sections and then, to reduce the theoretical uncertainty,
by taking
into account the full NLO\scite{aversa} information on the partonic scattering
processes.

We plot in fig.~\ref{fig1} the LO cross sections,
differential in $\pt$ and
integrated over the
$\eta<|0.5|$ range,
for producing a $\jpsi$ via
fragmentation, either directly or after radiative decay of a $\chi$ state.
The values of the various parameters entering into the calculation are listed
in
Table~\ref{table1}, and we are using $\mu = \mu_0$ for the
factorization/renormalization (f/r) scales.
The curves labeled by $\chi$ are due to gluon
fragmentation only. We have not included $c\to\chi$ fragmentation
contributions since, from the total fragmentation probabilities listed
in~\cite{chen}, they can be predicted to lie two orders of magnitude
below the $c\to\jpsi$ curve and be therefore surely negligible.

 From inspection of fig.~\ref{fig1}
the contributions from $\chi_1$ and $\chi_2$ states can
be clearly seen to dominate all over the $\pt$ range considered.

\begin{table}
\small
\begin{center}
\begin{tabular}{|l|r|c|}
\hline
$\as(2m_c)$ & 0.26& \\
\cline{1-2}
$m_c$ & 1.5 GeV & Ref.~\cite{bygluon}\\
\cline{1-2}
$|R_0(0)|^2$ & $(0.8$ GeV)$^3$ &\\
\hline
$H_1$ & 15 MeV &\\
\cline{1-2}
$H_8'(m_c)$ & 3 MeV & Ref.~\cite{bychi} \\
\hline
BR($\chi_0\to\jpsi$) & 0.007&\\
\cline{1-2}
BR($\chi_1\to\jpsi$) & 0.027& Ref.~\cite{pdg}\\
\cline{1-2}
BR($\chi_2\to\jpsi$) & 0.014& \\
\cline{1-2}
BR($\jpsi\to\mu^+\mu^-$) & 0.0597& \\
\hline
\begin{tabular}{l}
Initial scale for\\[-5pt]
charm fragm. evolution ($\mu_{ini}^c$)
\end{tabular} & 3$m_c$ & Ref.~\cite{bcycharm}\\
\hline
\begin{tabular}{l}
Initial scale for\\[-5pt]
gluon fragm. evolution ($\mu_{ini}^g$)
\end{tabular} & 2$m_c$  & Ref.~\cite{bygluon}\\
\hline
\begin{tabular}{l}
Initial scale for\\[-5pt]
induced gluon evol. ($\mu_{ini}^{ind}$)
\end{tabular} & 4$m_c$  & Ref.~\cite{cyind}\\
\hline
\end{tabular}
\parbox{10cm}{
\caption{
\label{table1}
\small Summary of the parameters. We have used for
the fragmentation functions the same values given
in the cited references, and the branching ratios as quoted by the Particle
Data Group.}
}
\end{center}
\end{table}

Next we compare, in fig.~\ref{fig2}, the results obtained for the dominant
$\chi_1 + \chi_2$ contribution in the LO approach with those obtained
by inserting also the
next-to-leading (NLO) partonic cross sections, to order $\as^3$,
with $\as$ evaluated to  two loop accuracy. We
remark that the latter does not result in a full NLO calculation as the
initial state evaluation and the evolution kernels of the fragmentation
functions
are still at leading order. We have explicitly checked that the inclusion
of NLO evolution kernels does not change appreciably our results.
Figure~\ref{fig2} clearly shows that the higher order
terms enhance the cross section by a factor about 1.5. This is consistent
with previous studies of higher order corrections in
heavy quark\scite{NDE,wim}
and inclusive jets\scite{aversa,ellis} production in hadron collisions.
The effect of variations of the f/r
scales $\mu$ between $0.6\>\mu_0$ and $2\mu_0$ is also shown. As expected,
the inclusion of the NLO terms reduces the sensitivity to scale variations.

One more theoretical uncertainty has to be considered, namely the freedom to
choose the initial scale for the fragmentation functions evolution around
$M_\jpsi$. Figure~\ref{fig2bis} shows the effect of varying $\mu_{ini}^g$ in
the range 2-4~GeV.

Finally we show, in figure~\ref{fig3} and in table~\ref{table2},
our final prediction for $\jpsi$
production by adding the mechanism of fragmentation to the direct
one\scite{mlmpriv}
and to the
production from $B$ decays as taken from ref.~\cite{mlm}, together with the
reported theoretical uncertainty. The bands are made
by choosing the highest and lowest curve which could be obtained by varying
some of the parameters: the factorization/renormalization scale
and the value of $\Lambda$ in the work of
ref.~\cite{mlm,mlmpriv}, again the same
scale and
the initial scale $\mu_{ini}^g$ in our result. The values of the parameters
used
for each curve are given in the caption. The total result is obtained by adding
togheter the two highest and the two lowest curves respectively. The size of
the
the fragmentation contribution is seen to be comparable with the previous
estimate for the sum of the two mechanisms considered up to now, leading
therefore to a sizeable enhancement of the predicted overall production rate,
which we also show in the figure.

\begin{table}
\small
\begin{center}
\begin{tabular}{|l||l|l||l|l||l|l|}
\hline
&\multicolumn{2}{|c||}{direct\cite{mlmpriv} + $B$ decays\cite{mlm}}&
\multicolumn{2}{|c||}{fragmentation}
&\multicolumn{2}{|c|}{total}\\
\hline
$\pt$&lower & upper & lower & upper & lower & upper \\
\hline
\hline
5.5  &$.52\times 10^0$&$.10\times 10^1$&$.12\times 10^1$&$.28\times 10^1$&
$.17\times10^1$&$.39\times 10^1$\\
\hline
7.5  &$.17\times 10^0$&$.36\times 10^0$&$.30\times 10^0$&$.75\times 10^0$&
$.46\times 10^0$&$.11\times 10^1$\\
\hline
9.5 &$.68\times 10^{-1}$&$.14\times 10^0$&$.98\times 10^{-1}$&$.25\times 10^0$&
$.17\times 10^0$&$.39\times 10^0$\\
\hline
11.5 &$.29\times 10^{-1}$&$.67\times 10^{-1}$&$.39\times 10^{-1}$&
$.10\times 10^0$&$.67\times 10^{-1}$&$.17\times 10^0$\\
\hline
13.5 &$.14\times 10^{-1}$&$.31\times 10^{-1}$&$.17\times 10^{-1}$&
$.47\times 10^{-1}$&$.32\times 10^{-1}$&$.79\times 10^{-1}$\\
\hline
15.5 &$.77\times 10^{-2}$&$.18\times 10^{-1}$&$.86\times 10^{-2}$&
$.24\times 10^{-1}$&$.16\times 10^{-1}$&$.41\times 10^{-1}$\\
\hline
17.5 &$.43\times 10^{-2}$&$.99\times 10^{-2}$&$.46\times 10^{-2}$&
$.13\times 10^{-1}$&$.89\times 10^{-2}$&$.23\times 10^{-1}$\\
\hline
19.5 &$.25\times 10^{-2}$&$.62\times 10^{-2}$&$.26\times 10^{-2}$&
$.73\times 10^{-2}$&$.52\times 10^{-2}$&$.13\times 10^{-1}$\\
\hline
\end{tabular}
\parbox{10cm}{
\caption{
\label{table2}
\small Numerical values of the cross sections (nb/GeV) plotted in
fig.~\protect\ref{fig3} (BR($\jpsi\to\mu^+\mu^-$) included).}}
\end{center}
\end{table}

When we finally compare with CDF data points\scite{cdf} we see that they are
now more compatible with the
theoretical band. This improves sensibly the previous situation, where
only by making very
extreme choices of the parameters one could get close to the experimental
findings.

To conclude, we have considered the inclusive production of $\jpsi$ in hadron
collisions in the framework of fragmentation functions. We have shown
explicitly
that the production and successive radiative decays of the $\chi$ states plays
a dominant role. The overall theoretical estimate, including the contribution
from $B$ decays, is nearly
consistent with the experimental observations.

A more detailed analysis, and a comparison with the most recent Tevatron data,
will be presented elsewhere\cite{mm}.

\vspace{.2cm}
One of us (M.C.) would like to thank Stefano Moretti, Marzia Nardi and Roberto
Tateo for useful suggestions.

\begin{figure}[h]
\begin{center}
\parbox{10cm}{
\caption{
\label{fig1}
\small Leading order differential cross sections due to various
fragmentation processes:
$c$, $g$, induced gluon fragmentation to $\jpsi$ and gluon fragmentation
to $\chi$ followed by radiative decay to $\jpsi$ are shown. Parameters as in
table~\protect\ref{table1}.}
}
\end{center}
\end{figure}

\begin{figure}[h]
\begin{center}
\parbox{10cm}{
\caption{
\label{fig2}
\small Leading vs. next-to-leading cross section for producing a $\jpsi$ via
fragmentation. Only the dominant $\chi_1$ and $\chi_2$ contributions are
included.}
}
\end{center}
\end{figure}

\begin{figure}[h]
\begin{center}
\parbox{10cm}{
\caption{
\label{fig2bis}
\small Theoretical uncertainty due to changes in the initial scale
$\mu_{ini}^g$ for the fragmentation functions evolution.}
}
\end{center}
\end{figure}

\begin{figure}[h]
\begin{center}
\parbox{10cm}{
\caption{
\label{fig3}
\small Theoretical prediction for $\jpsi$ production at Tevatron, compared to
experimental data from CDF\protect\scite{cdf}. Both the old result (dotted
line)
from
ref.~\protect\cite{mlm,mlmpriv}
 (upper curve: $\mu = m_T/4,~\Lambda = 275$~MeV; lower
curve $\mu = m_T,~\Lambda = 215$~MeV) and the new fragmentation (dashed line)
contribution (upper curve:
$\mu = 0.6\mu_0$, $\mu_{ini}^g = 4$~GeV; lower curve: $\mu = 2\mu_0$,
$\mu_{ini}^g = 2$~GeV) are included.}
}
\end{center}
\end{figure}


\end{document}